\documentclass[conference]{IEEEtran}
\usepackage[utf8]{inputenc}
\usepackage{graphicx}
\usepackage{float}
\usepackage{color, colortbl}
\usepackage{xcolor}
\usepackage{array}
\usepackage{multirow}
\usepackage{footnote}
\usepackage{cite}
\usepackage{threeparttable}

\IEEEoverridecommandlockouts

\makesavenoteenv{tabular}

\RequirePackage[normalem]{ulem} 
\RequirePackage{color}\definecolor{RED}{rgb}{1,0,0}\definecolor{BLUE}{rgb}{0,0,1} 

\begin{document}
 \title{Localization Procedure for Randomly Deployed WSNs based on the Composability of Position Estimation Protocols}

  \author{
      \IEEEauthorblockN{Luis Sanabria-Russo\IEEEauthorrefmark{0}, Cristina Cano\IEEEauthorrefmark{0}, Boris Bellalta\IEEEauthorrefmark{0}}
      \IEEEauthorblockA{\IEEEauthorrefmark{0}Universitat Pompeu Fabra, Barcelona, Spain
      \\\{luis.sanabria, cristina.cano, boris.bellalta\}@upf.edu}
  }


\IEEEoverridecommandlockouts
\IEEEpubid{\makebox[\columnwidth]{978-1-4673-2480-9/13/\$31.00 ~\copyright~2013 IEEE \hfill} \hspace{\columnsep}\makebox[\columnwidth]{ }}
\maketitle

\begin{abstract}
\boldmath Wireless Sensor Networks (WSNs) are composed of nodes that gather metrics such as temperature, pollution or pressure from events generated by external entities. Localization in WSNs is paramount, given that the collected metrics must be related to the place of occurrence. This work presents an alternative way towards localization in randomly deployed WSNs based on the combination of individual position estimation protocols (\emph{composability}). Our solution provides a flexible approach towards the localization problem in WSNs, which by means of composability it is able to locate more nodes than exclusively using a single localization protocol, while maintaining the same low levels of battery consumption.

\end{abstract}


%
%

\section{Introduction} \label{introduction}
  \IEEEPARstart{L}{ocalization} in randomly deployed WSNs has been a focus of interest in the research community. Its characteristics, like ease of deployment, suppose important advantages for some type of applications (nodes can be air-dropped off airplanes~\cite{airDroppedVolvano}). Global Positioning Systems (GPS) had been used to locate each node in the network. Nevertheless, because of the tightly constrained power source equipped in these nodes (normally two AA batteries) reducing the number of GPS modules is a viable way to increase the network lifetime while decreasing the budget.

To spread the implementation of this type of networks, localization protocols try to take the most out of the extremely constrained resources available. Limited battery, constrained processing power, constrained form-factor and cost are some of the limitations faced by each node~\cite{AkyildizWSNs}.

Localization protocols are often divided into two categories, called range-based and range-free. The former makes use of ranging techniques like Received Signal Strength Indicator (RSSI) in order to make straight-line distance estimations between the not-located nodes (called \emph{unknown}) and a reference node (called \emph{Anchor}) which broadcasts its location information in a packet type called \emph{Beacon}. The latter category just performs position estimations based on the effective connections among nodes. 

In some cases, one category might be more suitable than the other. For example, applications requiring coarse accuracy and running for very long periods of time might only need the simplicity offered by some range-free localization protocols. On the other hand, high-accuracy-demanding applications ask for localization protocols able to comply with strict accuracy requirements which are often achieved by combining several ranging techniques.

Although there are numerous protocols, none has proved to outperform the others under all possible scenarios and conditions; in \cite{composability} the authors combine and coordinate the execution of different position estimation protocols (from here on referred to as \emph{composability} of localization protocols) in order to leverage the weaknesses of some protocols with the strengths of others. Their proposal proved to be effective and capable of locating 100\% of the nodes in the deployment. Nevertheless, there are no considerations regarding the impact on battery consumption, estimation error and order of protocol execution.


This work extends the contribution of~\cite{composability} by addressing these issues and proposes a distributed and adaptive localization procedure for randomly deployed WSNs. This is achieved by having a clear understanding of the selected localization protocols' best-working conditions and network deployment considerations. A localization protocol is found suitable when, while complying with the deployment considerations, its best-working conditions are also reached. By implementing composability, the localization procedure is able to increase the number of located nodes compared to the individual execution of localization protocols.

The rest of the paper is organized as follows: a short literature review is presented in Section~\ref{literature} and in Section~\ref{locProc} the proposed localization procedure is described. A comparison between results gathered from the individual execution of two example localization protocols and those derived by the localization procedure are shown in Section~\ref{simulation}. Conclusions are drawn in Section~\ref{conclusions}.

\section{Literature Review} \label{literature}
  Range-based and range-free localization protocols use different set of techniques in order to estimate the position of an \emph{unknown} node~\cite{rang:loc:techniques}. Range-based localization protocols gather information about the received signal strength as indicator of range towards the transmitter. Ranging techniques are often combined with localization techniques like trilateration and multilateration to derive a point where the \emph{unknown} node probably lies. On the other hand, range-free localization protocols use the effective connections, usually of the type \emph{unknown-Anchor}, to draw a plane that represents the intersection of the coverage areas of such \emph{Anchors}. This area is composed of all the possible points where the \emph{unknown} node is probably located.

In this section, some well-used ranging and localization techniques are reviewed.

\subsection{RSSI ranging technique} \label{rssi}
Commercially available nodes, like the Crossbow TelosB~\cite{telosB}, are capable of reporting RSSI measures. This metric is related to the received signal strength at the node and although it is heavily affected by channel uncertainties (like shadowing and multi-path), it can be used to make rough range estimations~\cite{rang:loc:techniques}.

Ranging techniques incur in additional battery consumption since multiple Beacon readings should be performed in order to reduce ranging errors; which requires an increased channel listening time.

\subsection{Trilateration and multilateration localization techniques} \label{lateration}
Range-based localization protocols use range measurements as input to more complex localization techniques. Trilateration places the \emph{unknown} node $j$ at the edge of a circumference of radius $d_{ij}$, where $i$ is usually an \emph{Anchor} placed at the center of the circumference. When three \emph{Anchors} ($i=1,2,3$) are connected to node $j$, the intersection of these circumferences results in the position of the node.

Multilateration also uses range measurements, quite differently this technique consists on minimizing a set of $n$ equations ($i=1,2,3,...,n$) as shown in~(\ref{eq:lateration}).

\begin{equation}\label{eq:lateration}
 f_{i}(x_{j},y_{j})=d_{ij}-\sqrt{(x_{i}-x_{j})^2+(y_{i}-y_{j})^2}
\end{equation}

In~(\ref{eq:lateration}), $(x_{i},y_{i})$ are \emph{Anchor} $i$'s coordinates and $(x_{j},y_{j})$ represents the \emph{unknown} node's estimated position~\cite{rang:loc:techniques}.

These localization techniques rely on exact distance measurements and the resulting error is directly related to the ranging technique used. That is, although trilateration's mathematical solution is a point on a plane, the estimation carries an underlying error resulting from inexact range measurements. As it also happens with multilateration~\cite{AkyildizWSNs}.

Furthermore, localization techniques like Lateration incur in additional battery consumption mostly related to the minimization of a set of equations like~(\ref{eq:lateration}) and the ranging technique used. As mentioned in~\cite{laterationSpecs}, this additional energy consumption with RSSI ranging technique and four \emph{Anchors} is of around~$1.961$~mJ per execution of the Lateration algorithm.

\subsection{Bounding-Box}
This range-free protocol consists on placing the \emph{unknown} node at the intersection of the coverage areas generated by the surrounding \emph{Anchors}. The resulting intersection is usually called Location Area (LA). 


Because the \emph{unknown} node does not need to perform ranging measurements, this technique incurs in a reduced energy consumption when compared to other range-based protocols, like Lateration.

Optionally, the LA can be further reduced defining constraints based on ranging techniques; like angle of arrival or considering variable coverage areas. Both approaches are proposed and implemented in~\cite{convexEstimation}. 

\subsection{Composability of localization protocols}
The approach proposed by the authors of~\cite{composability} is based on the observation than current protocols either make simplifying assumptions (Line of Sight (LoS) scenarios, exact measurements, high \emph{Anchor} density, known distribution of the nodes) or require sophisticated hardware (like in the case of Angle of Arrival (AoA) or the tight synchronization needed in Time of Arrival (ToA) Ranging Techniques~\cite{AkyildizWSNs}). They also argue that localization protocols that do not make these assumptions provide greatly inaccurate results.

Their approach consists in storing multiple localization protocols in every node. Then, these protocols are executed according to a predefined sequence triggered by accuracy thresholds. 

%

Although this approach succeeded at combining different localization protocols, there is lack of detailed information regarding which protocols are to be executed first and why. Also, its impact on network lifetime is left as a future research topic.

\section{Localization Procedure} \label{locProc}
  The composability of localization protocols proposed in~\cite{composability} tries to leverage the weaknesses that some protocols may have under certain conditions. Nevertheless, there might be opportunities where the predefined sequence of protocol execution would result in increased errors due to lack of consideration of the \emph{unknown} node's network-environment or the priorities of the deployment.

We propose a localization procedure which focuses on considering the protocols' best-working environmental conditions and the WSN deployment considerations in order to make the most beneficial protocol selection instead of a static sequential execution.


\begin{figure}[tb]
  \centering
  \includegraphics[width=\linewidth]{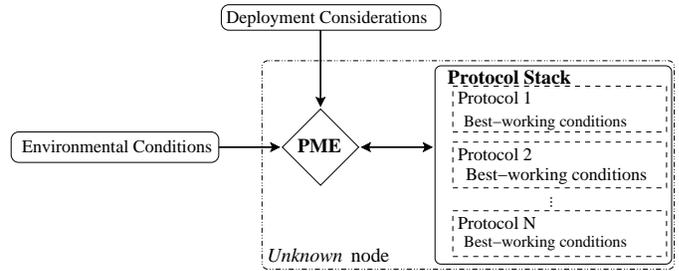}
  \caption{Localization procedure: architecture
  \label{fig:LocProc}}
\end{figure}

\subsection{Best-working environmental conditions}\label{bestWorkingConditions}
Some protocols perform better than others under different conditions. For instance, some \emph{Anchor}-based localization techniques (like the ones described in Section~\ref{literature}) require more connections to \emph{Anchors} than others~\cite{rang:loc:techniques}.

When referring to best-working environmental conditions, it is to point out network-environment metrics that would help a determined localization protocol to work more efficiently, like: number of effective connections of the type \emph{unknown-Anchor}, current delay, available bandwidth, network size or processing capabilities of the nodes.

Up-to-date information of the node's environmental conditions aids the process of determining which localization protocol is more capable of achieving the deployment considerations.

\subsection{Deployment considerations}\label{deploymentConsiderations}
Each deployment has defined goals and restrictions, like: long/coarse network lifetime, high/coarse accuracy, short/coarse localization traffic overhead or high/low localization protocol convergence time. These are tightly related to the application running over it.

Each localization protocol has its best-working environmental conditions, that when complied allow the protocol to provide satisfactory results and follow the deployment considerations.

\subsection{Pattern Matching Engine (PME)}\label{PME}
The PME is a module inside the localization procedure responsible for translating the \emph{unknown} node's environmental conditions into localization protocols than could comply with the deployment considerations. That is, for certain deployment considerations the PME will select a set of appropriate localization protocols where their best-working environmental conditions are met. If all the conditions are satisfied, the PME prioritizes the protocol that better complies with the deployment considerations. 

The architecture of the localization procedure is shown in Fig.~\ref{fig:LocProc}, where it can be appreciated the PME gathering environmental conditions and deployment considerations to select the appropriate localization protocol.


\section{Evaluation} \label{simulation}
  This work considers two well-known distributed localization protocols for testing the proposed localization procedure: Lateration and Bounding-Box. Some of their differences are highlighted in Table~\ref{table:protocols}.

\begin{table}[tb]
  \centering
  \begin{threeparttable}[t]
    \caption{Localization Protocols' characteristics}
    \label{table:protocols}
    \begin{tabular}{c||c||c}
    \hline
    \bfseries Characteristic & \bfseries Lateration & \bfseries Bounding-Box\\
    \hline\hline 
    Env. Conditions & At least 4 \emph{Anchors} & At least 1 \emph{Anchor}\\
    Accuracy & 2-10 meters & Coarse\tnote{1}\\
    Energy Consumption & Low~\cite{laterationSpecs} & Very low\tnote{2}\\
    \hline
    \end{tabular}
    \begin{tablenotes}
    \item [1] Location area upper-bounded by \emph{Anchor}'s radio range (R).
    \item [2] Can be treated as a discrete problem.
    \end{tablenotes}
  \end{threeparttable}
\end{table}

\begin{figure}[tb]
  \centering
  \includegraphics[width=0.7\linewidth, angle = -90]{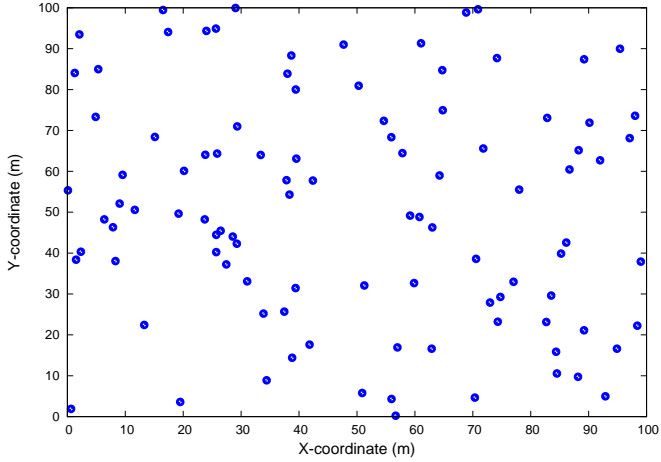}
  \caption{Example random deployment of nodes
  \label{fig:topology}}
\end{figure}

In order to reveal the impact of the proposed localization procedure in terms of battery consumption, number of located nodes and localization error; a thousand simulations are performed per \emph{Anchor} density (from 10\% up to 100\% at 10\% increments) using a customized extension of the SENSE network simulator~\cite{sense}. The hardware and Medium Access (MAC) layer parameters implemented are presented in Table~\ref{tab:MAC_param}. Two propagation models are used: free space and a time-invariant and symmetrical shadowing model (from here on: Free space and Shadowing models respectively). The characteristics of the testing plane are highlighted in Table~\ref{tab:testingPlanes}. Nodes are randomly and uniformly distributed over the testing plane (as in Figure~\ref{fig:topology}) and the position estimation is based only on the received location information from Beacons. This is done to evaluate different \emph{Anchor} densities against a single \emph{unknown} node (i.e. 100\% \emph{Anchor} density means 
that a determined node will receive Beacons from all its neighbors and use the received location information to estimate its position, regardless if the recipient is an \emph{Anchor}).

To contrast the behavior of the localization procedure against the individual execution of the proposed localization protocols, the deployment considerations are set accordingly with the capabilities of the protocols (large network lifetime and coarse accuracy). The PME deterministically selects the appropriate protocols as it is explained in Section~\ref{PME}.


Results are shown with 99\% confidence intervals.

\begin{table}[tb]
  \begin{threeparttable}[t]
    \caption{Hardware and CSMA/CA Parameters}
    \label{tab:MAC_param}
    \begin{tabular}{c||c||c}
    \hline
    \bfseries Component & \bfseries Parameter & \bfseries Value\\
    \hline\hline 
    \multirow{8}{*}{Hardware} & Data rate & $19.2$~kbps\\ 
			      & TX power & $0$~dBm\\ 
			      & Reception threshold & $-148$~dBm\\ 
			      & Carrier sense threshold & $-148$~dBm\\ 
			      & Power consumption in TX mode & $24.75$~mW\\ 
			      & Power consumption in RX/idle mode & $13.5$~mW\\ 
			      & Power consumption in sleep mode & $15~\mu\rm{W}$\\ 
    \hline
    \multirow{4}{*}{CSMA/CA} & Headers & $11$~bytes\\ 
			      & Beacon size & $40$~bytes\\ 
			      & Contention window & $128$\\ 
			      & Slot time & $417~\mu\rm{s}$\\ 
    \hline
    \end{tabular}
  \end{threeparttable}
\end{table}


\begin{table}[tb]
  \centering
  \begin{threeparttable}[t]
  \caption{Characteristics of the testing plane based on~\cite{convexEstimation,fieldDimmensions}}
  \label{tab:testingPlanes}
  \begin{tabular}{c||c}
  \hline
  \bfseries Characteristic & \bfseries Value\\
  \hline\hline
  Area & $100\times100$~m$^{2}$\\
  Surface & Flat\\
  Distribution of nodes & Uniformly random\\
  \hline
  \end{tabular}
  \end{threeparttable} 
\end{table}

Each time a node connects with a new \emph{Anchor} (effectively receives its Beacons), the PME decides which localization protocol to execute. In the proposed simulation (and following Table~\ref{table:protocols}) if more than three \emph{Anchors} are connected to the \emph{unknown} node, then the PME will execute Lateration, otherwise Bounding-Box is selected. Further connections will lead the PME to reevaluate the node's situation and sequentially execute the appropriate localization protocol. As for Lateration, this can go on until six \emph{Anchors} are connected. Beyond this number, the accuracy refinements are not as significant to justify the penalty in battery consumption~\cite{beaconLimits}.

\subsection{The effect of the tested channel models on the number of connection to \emph{Anchors}}\label{channel_considerations}
\begin{figure}[tb]
  \centering
  \includegraphics[width=0.7\linewidth, angle = -90]{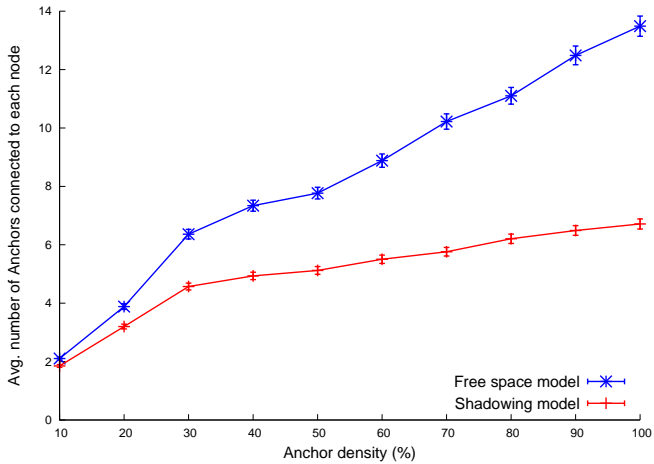}
  \caption{Average number of \emph{Anchors} connected to each node}
  \label{fig:channelAndBeacons}
\end{figure}

In Figure~\ref{fig:channelAndBeacons}, it is appreciated how the different propagation models affect the reception of Beacons; which in this evaluation is the sole environmental condition considered by the PME.

Nodes in the Shadowing model are prone to more collisions than those in the Free space model. When there is high concentration of neighboring \emph{Anchors}, it is more probable that collisions occur. This results in a decreased number of \emph{Anchors} connected to each node, which has a direct impact on battery life, accuracy and number of located nodes.

\subsection{Individual execution of Lateration and Bounding-Box localization protocols}\label{individual_execution}
In order to better analyze the impact of the localization procedure on the network, metrics are gathered from the individual execution of the tested localization protocols.

These metrics reflect the protocols' impact on battery consumption, number of located nodes and the position estimation error.

\subsubsection{Battery consumption}\label{individual_battery_consumption}
\begin{figure}[tb]
  \centering
  \includegraphics[width=0.7\linewidth, angle = -90]{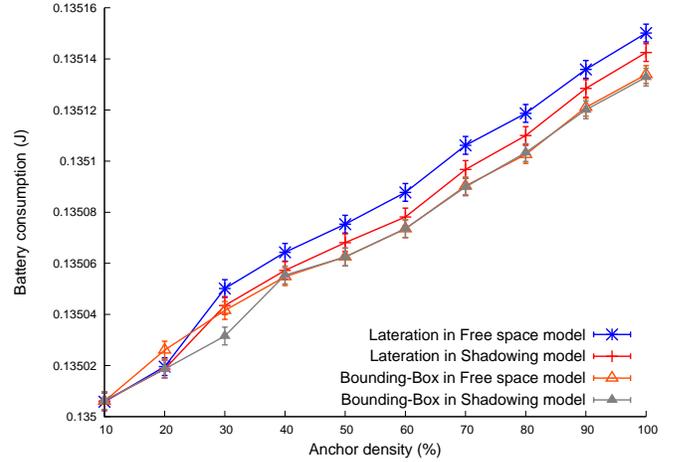}
  \caption{Battery consumption of the individual execution of Lateration and Bounding-Box
  \label{fig:battery}}
\end{figure}

apart from the battery consumption related with the normal operation of the nodes (listening the channel and Beacon reception), Lateration has an additional battery consumption associated with the execution of the algorithm (as mentioned in Section~\ref{lateration}). This seems to be increased in the Free space model because of the greater average number of \emph{Anchors} connected to each node (see Figure~\ref{fig:channelAndBeacons}~and~\ref{fig:battery}).

In the case of Bounding-Box, there is not additional battery consumption related to the execution of this algorithm. This is the reason why its added battery consumption is considered negligible when compared to Lateration.


\subsubsection{Located nodes}
\begin{figure}[tb]
  \centering
  \includegraphics[width=0.7\linewidth, angle = -90]{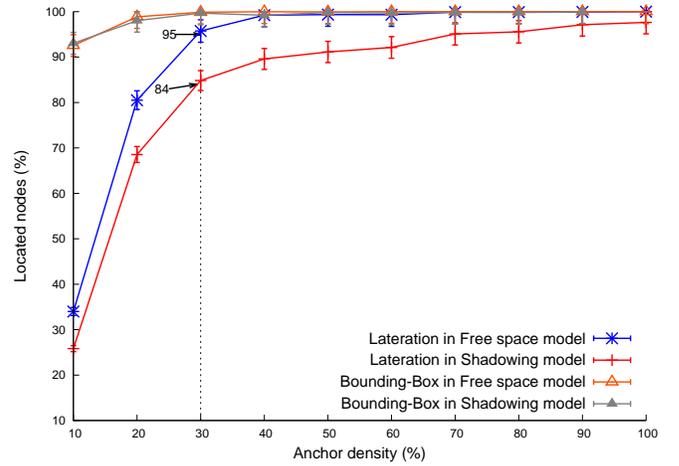}
  \caption{Number of located nodes per tested localization protocol
  \label{fig:locNodes}}
\end{figure}

these are the nodes that successfully execute either of the localization protocols, resulting in a location estimation (see Figure~\ref{fig:locNodes}).

In the case of Lateration, at 30\% \emph{Anchor} density around 84\% and 95\% of the nodes get located in the Free space and Shadowing models respectively. Lower numbers are appreciated at 10-20\% \emph{Anchor} density due to the reduced/inexistent Beacons received at these densities.

Bounding-Box shows higher number of located nodes at 30\% \emph{Anchor} density (nearly 99\% in both propagation models) mainly due to a more coarse restriction for the execution of this protocol (only one Beacon).

\subsubsection{Error}
\begin{figure}[tb]
  \centering
  \includegraphics[width=0.7\linewidth, angle = -90]{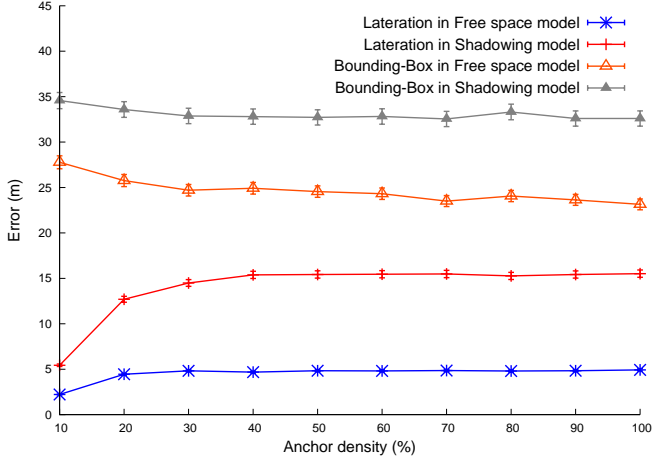}
  \caption{Straight line error from the estimated to the real node's position
  \label{fig:error}}
\end{figure}

the proposed measure of error only considers nodes that were able to execute either of the localization protocols. It is defined as the straight line distance (in meters) between the node's estimated location and its real position (see Figure~\ref{fig:error}).

The error in Lateration is related to inexact ranging measurements, which is greater in the Shadowing model given that for the calculations the node always assumes a Free space model. The location accuracy does not seem to improve significantly with the connection of more than six \emph{Anchors} without sacrificing battery life~\cite{beaconLimits}. Furthermore, it worsens with the degraded channel conditions imposed by the Shadowing propagation model.

For Bounding-Box, the Shadowing model reduces the average number of \emph{Anchors} received at the \emph{unknown} node, which translates in the elimination of some of the constraints that allow this protocol to increase its accuracy.


\subsection{Localization procedure execution}\label{locProc_executions}
As mentioned in Section~\ref{PME}, PME will pick a localization protocol that given the node's environmental conditions, could comply with the deployment considerations.

\subsubsection{Battery consumption}
\begin{figure}[tb]
  \centering
  \includegraphics[width=0.7\linewidth, angle = -90]{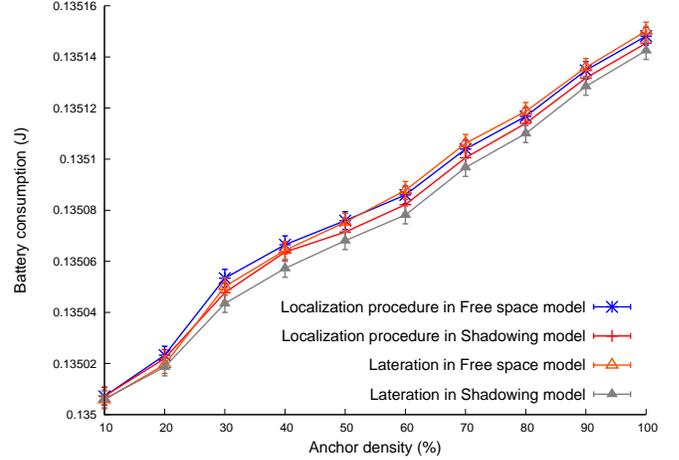}
  \caption{Localization procedure's associated battery consumption when compared with Lateration
  \label{pme:battery}}
\end{figure}

the difference between the battery consumption associated with the proposed localization procedure and that of Lateration is very small. For this reason they are considered similar (see Figure~\ref{pme:battery}).

Bounding-Box adds negligible battery consumption (as mentioned in Section~\ref{individual_battery_consumption}), therefore it is not included in Figure~\ref{pme:battery}, which only attempts to compare the average battery consumption of the individual execution of Lateration and the amount consumed by the proposed localization procedure.

\subsubsection{Located nodes}\label{locProc_locatedNodes}
\begin{figure}[tb]
  \centering
  \includegraphics[width=0.7\linewidth, angle = -90]{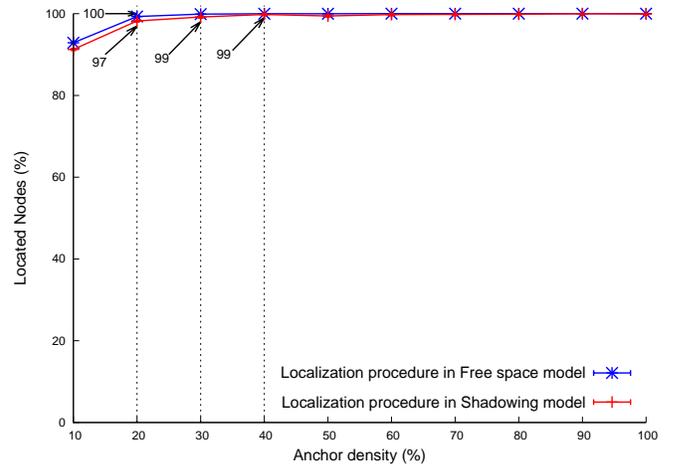}
  \caption{Located nodes with the proposed localization procedure
  \label{pme:locNodes}}
\end{figure}

the sum of located nodes (either with Lateration or Bounding-Box) reaches 99\% at \emph{Anchor} densities around 30\% (see Figure~\ref{pme:locNodes}).

With Free space model 100\% localization is achieved at 20\% \emph{Anchor} density. On the other hand, in the Shadowing model 100\% localization is achieved at slightly higher densities, on average around 40\%.

The number of located nodes with the proposed localization procedure exceeds those of Lateration, in fact Figure~\ref{pme:locNodes} looks more like the curves of Bounding-Box in Figure~\ref{fig:locNodes}.

\subsubsection{Error}
\begin{figure}[tb]
  \centering
  \includegraphics[width=0.7\linewidth, angle = -90]{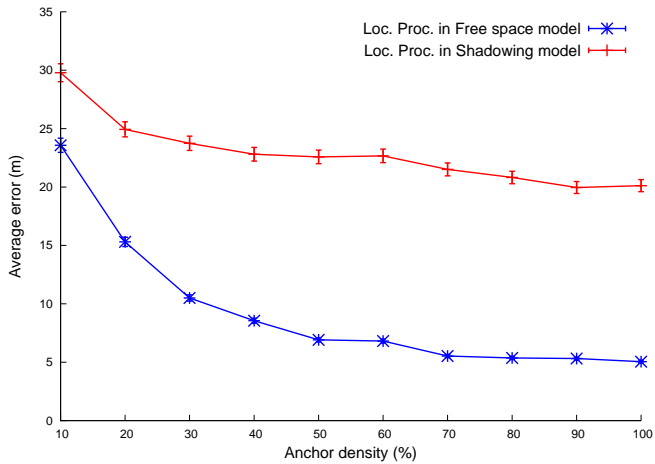}
  \caption{Associated error after protocol selection
  \label{pme:error}}
\end{figure}

This measure illustrates the average distance in meters between each node's estimated location and its real position. The prefix \emph{Loc. Proc.} in Figure~\ref{pme:error} highlights the fact that these are results gathered from the execution of the localization procedure.

Figure~\ref{pme:error} displays the average error for each type of channel model. It also considers the number of nodes executing either Lateration or Bounding-Box. So, the average error is expressed as: $Avg_{A}^{ch} = (E_{L}~n_{L} + E_{BB}~n_{BB}) /n_{L}+n_{BB}$, where $Avg_{A}^{ch}$ is the average error when a certain channel model ($ch$) is used and the density of \emph{Anchor} nodes is $A$. The parameter $n_{L}$ refers to the number of nodes executing Lateration while $n_{BB}$ are the same but in the case of Bounding-Box. $E_{L}$ and $E_{BB}$ refer to the average line error for nodes executing Lateration and Bounding-Box respectively.

Due to increased ranging measurement errors, the estimation in the Shadowing model presents more erroneous estimations than the Free space model, as in Figure~\ref{fig:error}. Although at $30$\% \emph{Anchor} density the localization procedure incurs in greater average error as compared with the individual execution of Lateration, it manages to increase the average number of located nodes. This is significantly important, given that without the localization procedure many nodes were to be left without a location estimation, or what it is the same as having an undetermined measure of error.

A carefully selected set of localization protocols working with different ranges of environmental conditions, ensures that most of the nodes in the deployment get located. In the testings presented in this section, Bounding-Box is the responsible for locating the most isolated nodes, while Lateration focuses on accuracy. Selecting and characterizing more accurate protocols will reduce the errors and maintain the high number of located nodes that the localization procedure achieves. All of this while preserving the levels of battery consumption similar to the individual execution of the selected protocols.

\section{Conclusions} \label{conclusions}

In this work a new and flexible approach to the localization problem in randomly-deployed WSNs is presented. It extends the proposal of~\cite{composability}, which considers the composability of localization protocols as a robust solution. 

The localization procedure incorporates flexibility on the selection of localization protocols by determining which is more capable of achieving predetermined deployment considerations under the environmental conditions surrounding each \emph{unknown} node. Furthermore, it is designed to admit several localization protocols, definitions of environmental conditions and deployment considerations; which makes it a good choice for random deployments, like~\cite{airDroppedVolvano}.

A set of evaluations were preformed with two well-know localization protocols, referred to as Lateration (range-based) and Bounding-Box (range-free). Results show that the localization procedure is able to locate more nodes than by their individual execution, suggesting a more intelligent and flexible localization scheme that considers the current state of the nodes before making decisions about its future state.


In order to improve the current proposal, it is important to identify the environment metrics that correlate with the performance of each localization protocol to be used. Once understood, simple adjustments in the PME would enable it to comply with the deployment considerations in a more effective way. Moreover, the PME can be adapted to make a protocol selection based not only on its own, but also with the surrounding nodes' environmental conditions. This opens the door to more complex and centralized localization algorithms, like~\cite{pal2010localization}~and~\cite{alippi2006rssi}.
  
\section{Acknowledgements} \label{ack}
  This work has been partially supported by the Spanish Government, through the project CISNETS (Plan Nacional I+D+i, TEC2012-32354).

\bibliographystyle{Classes/IEEEtran}
\bibliography{IEEEabrv,ref}

\end{document}